\documentclass[preprint]{aastex}

\usepackage{psfig}

\newcommand\mzon   {M$_{\odot}$}
\newcommand\pp     {$\pm$}
\newcommand\pers     {s$^{-1}$}
\newcommand\micros  {$\mu$s}
\newcommand\Lunit {ergs s$^{-1}$}
\newcommand\funit {ergs s$^{-1}$ cm$^{-2}$}

\begin{document}

\righthead{The 2000 outburst light curve of SAX J1808.4--3658}
\slugcomment{}
\title{The erratic luminosity behavior of SAX J1808.4--3658 during
its 2000 outburst}

\author{Rudy Wijnands\altaffilmark{1,5}, Mariano
M\'endez\altaffilmark{2}, Craig Markwardt\altaffilmark{3}, Michiel van
der Klis\altaffilmark{4}, Deepto Chakrabarty\altaffilmark{1}, Ed
Morgan\altaffilmark{1}}

\altaffiltext{1}{Center for Space Research, Massachusetts Institute of
Technology, 77 Massachusetts Avenue, Cambridge, MA 02139-4307, USA;
rudy@space.mit.edu}

\altaffiltext{2}{SRON Laboratory for Space Research, Sorbonnelaan 2,
NL-3584 CA, Utrecht, The Netherlands}

\altaffiltext{3}{NASA/Goddard Space Flight Center, Code 662,
Greenbelt, MD 20771}

\altaffiltext{4}{Astronomical Institute ``Anton Pannekoek'',
       University of Amsterdam, Kruislaan 403, NL-1098 SJ Amsterdam,
       The Netherlands}

\altaffiltext{5}{Chandra Fellow}

\begin{abstract}
We report on the highly variable and erratic long-term X-ray
luminosity behavior of the only known accretion-driven millisecond
X-ray pulsar SAX J1808.4--3658 during its 2000 outburst, as observed
with the {\it Rossi X-ray Timing Explorer} ({\it RXTE}) satellite.
The maximum observed luminosity is $\sim2.5\times10^{35}$
\Lunit~(3--25 keV; for a distance of 2.5 kpc), which is approximately
a factor of ten lower than that observed during the 1996 and 1998
outbursts.  Due to solar constraints, the source could not be observed
for several months with {\it RXTE} before 21 January 2000. Therefore,
the exact moment of the outburst onset is unknown and the peak
luminosity could have been significantly higher.  On some occasions
SAX J1808.4--3658 was observed with luminosities of $\sim$10$^{35}$
\Lunit, but on other occasions it could not be detected with {\it
RXTE} resulting in typical upper limits of a few times $10^{33}$
\Lunit~(3--25 keV). The non-detections of the source during its 2000
outburst obtained with the {\it BeppoSAX} satellite demonstrate that
its luminosity was at times $<10^{32}$ \Lunit~(0.5--10 keV; Wijnands
et al. 2001). However, only a few days after these {\it BeppoSAX}
observations, we detected the source again with {\it RXTE} at high
luminosities, giving a factor of $>$1000 of luminosity swings in this
systems on time scales of days. The last detection of SAX
J1808.4--3658 with {\it RXTE} was on 2000 May 13, almost four months
after the first detection during this outburst. Due to the lack of
sensitivity and observations during the 1996 and 1998 outbursts, it
cannot be excluded that after those outbursts the source remained
active for months and that the source behavior during the 2000
outburst is not unique.  Long duration activity at low luminosities
has been observed in other transients (both neutron stars and black
holes), although not with such extreme variability which might point
to a different origin for this behavior for the millisecond X-ray
pulsar.

\end{abstract}


\section{Introduction \label{intro}}

Low-mass X-ray binaries (LMXBs) are binary systems containing a
compact object (either a neutron star or black hole) which is
accreting matter from a low-mass companion star ($<$1 \mzon).  A
particular sub-class of LMXBs, the X-ray transients, have received
special interest. Due to the high variations in accretion rate, these
systems can be studied under a very wide range of physical conditions.
They exhibit (sometimes recurrent) outbursts during which they
resemble in detail the persistent LMXBs. Most of the time, however,
these X-ray transient are in quiescence, during which no or hardly any
accretion onto the compact object takes place.

The outburst profiles of the X-ray transients are very diverse and
although the most recent versions of the disk instability model can
explain the outburst light curve properties in general, some important
issues remain unexplained (see Lasota 2001 for a review of disk
instability models). In most sources, the X-ray flux during the
outburst decreases steadily to quiescence levels, although some
transients exhibited a much more complex behavior (see, e.g., Chen,
Shrader, \& Livio 1997 and Bradt et al. 2000). Some sources exhibit
periods of low-level activity for several months to several years,
long after the outburst is over. Both black hole systems (e.g., 4U
1630--47 or XTE J0421+560: Kuulkers et al. 1997, Bradt et al. 2000;
Parmar et al. 2000) and neutron star systems (e.g., Aql X-1 and 4U
1608--52; Bradt et al. 2000) can exhibit this behavior.  Besides the
low-level of activity observed after a full outburst, Aql X-1 also
exhibited so-called ``failed'' outbursts, during which the source does
not reach typical outburst luminosity but low-level activity is
present for weeks (Bradt et al. 2000). The physical processes behind
these long episodes of low-activity are unclear. One possibility is
that these states indicate that the conditions for exhibiting
outbursts are marginal and when systems accrete close to the critical
value they might show low-level activity for several weeks to months
(see also Bradt et al. 2000).  In the remainder of this {\it Letter},
we refer to these long episode of low-level activity in X-ray
transients as ``the low-activity state''.

\subsection{SAX J1808.4--3658}

In September 1996, using the {\it BeppoSAX} Wide Field Cameras (WFC),
a new X-ray transient (designated SAX J1808.4--3658) was discovered
(in 't Zand et al.~1998). The neutron-star nature of the compact
object in this system was deduced from the observations of type-I
X-ray bursts from which also a distance of $\sim$2.5 kpc was derived
(in 't Zand et al. 1998; 2001).  The source had a maximum luminosity
of $\sim2.5\times10^{36}$ \Lunit~(3--25 keV) and could be detected for
several weeks. The source remained dormant until 9 April 1998, when it
was detected again, but this time with the proportional counter array
(PCA) aboard the {\it Rossi X-ray Timing Explorer} ({\it RXTE};
Marshall 1998).  Using the {\it RXTE}/PCA data, coherent 401 Hz
pulsations were discovered in the X-ray flux of SAX J1808.4--3658,
making this source the first known accretion-driven millisecond X-ray
pulsar (Wijnands \& van der Klis 1998a). It was also found that the
neutron star is in a 2 hr binary with a very low-mass companion star
(Chakrabarty \& Morgan 1998; see Bildsten \& Chakrabarty 2001 for a
discussion about the possible brown dwarf nature of the companion star
of SAX J1808.4--3658). During its 1998 outburst, its aperiodic rapid
X-ray variability (Wijnands \& van der Klis 1998b) and its X-ray
spectrum (Gilfanov et al. 1998; Heindl \& Smith 1998) were remarkably
similar to those of other LMXBs with similar luminosities.  Therefore,
it is very puzzling why persistent millisecond pulsations have so far
only been detected from this source and not from any other neutron
star LMXBs.

During its 1998 outburst, the source was detected for a few weeks
(see, e.g., Gilfanov et al. 1998 or Cui, Morgan, \& Titarchuk 1998 for
its 1998 light curve) before it returned to quiescence. SAX
J1808.4--3658 was detected for a third time on 21 January 2000 (again
with the {\it RXTE}/PCA), but this time at a flux level of about a
tenth of the fluxes observed during the September 1996 and April 1998
outbursts (van der Klis et al. 2000). When the source was detected,
the pulsations at 401 Hz were observed and during several {\it RXTE}
observations very strong violent flaring behavior with a repetition
frequency of about 1 Hz was present (van der Klis et al. 2000;
Wijnands et al. 2000). During this outburst, several {\it BeppoSAX}
observations were also performed to study the broad band spectrum of
the source.  Here, we report on the 2000 outburst light curve of SAX
J1808.4--3658 as obtained with the {\it RXTE}/PCA. The coherent
pulsations are reported by Morgan et al. (2001) and the {\it BeppoSAX}
observations by Wijnands et al. (2001).

\section{Observations and analysis}

After SAX J1808.4-3658 was first detected on 21 January 2000, the
source was frequently monitored with {\it RXTE} as part of our AO4 TOO
program on this source. A total of 41 observations were obtained for a
total of $\sim$150 ksec on source time.  During the observations, the
data were taken simultaneously in the Standard 1 (1 energy channel and
1/8 s time resolution) and Standard 2 modes (129 channels; 16 s
resolution), and the event mode E\_125US\_64M\_0\_1S (64 channels;
122\micros~resolution). The only observation during which different
modes were used was 40035-01-04-00G. After a brief interval of
non-detections of SAX J1808.4--3658, the source was found to be active
again and {\it RXTE} was slewed in real-time to the position of SAX
J1808.4--3658.  As a consequence, the high-timing modes used during
this observation were the same one as those used on the previous
target and were the GoodXenon1\_16s and GoodXenon2\_16s modes
(combined they have 256 channels for 2--60 keV and a time resolution
of $\sim$ 1 \micros). In this {\it Letter}, we only use the data
obtained with the standard modes. The results obtained with the high
time resolution modes are presented in Morgan et al. (2001).

The Standard 2 mode data were used to extract the count rates of the
source and the X-ray spectra. The count rates were only extracted for
proportional counter units (PCUs) 0 and 2, which were always on during
our observations. Spectra were extracted for all the detectors which
were on during the particular observations.  We used the tools
provided with ftools version 5.0.4 to extract the background data
(using the ``faint source'' model) which were used to correct the
count rates and the spectra.  We fitted the X-ray spectra between 3
and 25 keV (with a 1\% systematic error added). The spectra were
fitted with a power-law with a free-floating absorption column density
(the column densities obtained from the fit were consistent with zero
{\it and} with the Galactic value of $1.3\times10^{21}$ cm$^{-2}$
towards SAX J1808.4--3658; Dickey \& Lockman 1990). During several
observations, the power-law alone did not produce an acceptable fit to
the data and an extra low-energy ($<$5 keV) component needed to be
added. However, when fitting this component with a black-body, very
low $kT$ ($<$0.5 keV) values were obtained. Presently, it is unclear
whether this component is real or if it is due to residual calibration
uncertainties of the detectors at low energies. We will not discuss
this further in our paper. When no source flux could be detected we
obtained 95\% confidence upper limits to the flux by fixing the
power-law photon index to 2. The actual limits for the individual
observations (0.02--0.2 $\times10^{-10}$ \funit) depended upon the on
source time and the number of PCUs on during those observations.

In our analysis we also included the data obtained as part of {\it
RXTE}/PCA monitoring of the Galactic bulge region.  Observations began
in February 1999, and typically take place twice per week, except for
approximately 2.5 months between November and February when the sun is
near the Galactic center (Markwardt et al. 2000; Markwardt 2000).  The
data (in good-xenon mode) are obtained by scanning the spacecraft to
cover a large region.  Intensities of known sources are derived by
fitting a collimator model to the observed light curve, including the
contribution of a diffuse emission component (see Markwardt et
al. 2001 for details).  SAX J1808.4--3658 lies at the endpoint of one
of the scans, and because the spacecraft must dwell at that point
momentarily the exposure is 2--3 minutes.  The scans are thus
sensitive to variations at the $> 10^{-12}$ \funit~level.  Our TOO
observations were performed until 1 March 2000, but the bulge scan
observations continue.

\section{Results}

\begin{figure}[]
\begin{center}
\begin{tabular}{c}
\psfig{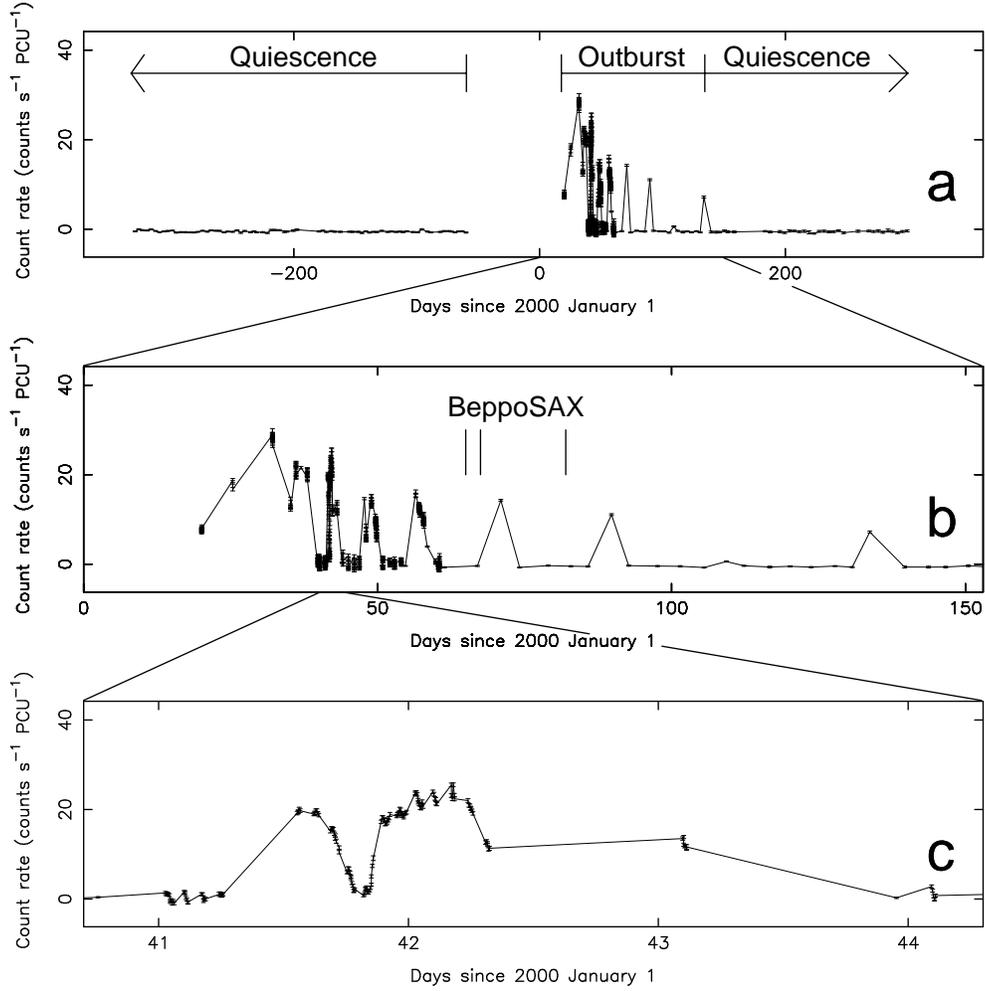}
\end{tabular}
\figcaption{{\it RXTE}/PCA 2--60 keV count rate curve of SAX
J1808.4--3658 during its 2000 outburst. In ({\it a}) the most
extensive curve (up to 26 October 2000) is shown, and in ({\it b}) and
({\it c}) close ups of the data are shown to show the strong
variability more clearly.  In ({\it b}), the vertical lines indicate
when the three {\it BeppoSAX} observations were taken (see Wijnands et
al. 2001). The count rates are background subtracted. The data points
obtained as part of the TOO proposal represent 256 s of data, whereas
those from the {\it RXTE}/PCA bulge scan data have a variable length
(typically 2--3 minutes). The gap in the data in ({\it a}) is due to
solar constraints, which did not allow {\it RXTE} observations of SAX
J1808.4--3658 in this period.
\label{fig:rxte_lc}}
\end{center}
\end{figure}

The 2--60 keV count rate curve of SAX J1808.4--3658 during its 2000
outburst is shown in Figure \ref{fig:rxte_lc}.  For display purposes,
we only use the data obtained with the {\it RXTE}/PCA bulge scan up to
26 October 2000. During the bulge scan observations performed after
that date, the source was not detected and only count rate upper
limits could be obtained. The gap in the data (between 2 November 1999
and 21 January 2000) in Figure~\ref{fig:rxte_lc}{\it a} is due to
solar constraints, which did not allow {\it RXTE} observations of SAX
J1808.4--3658 in this period.  Clearly, three episodes of distinct
behavior can be observed for this source. Before 2 November 1999, the
source was undetected with typical upper limits on the luminosity of a
few times $10^{33}$ \Lunit~(3--25 keV; hereafter, the quoted
luminosities are for this energy range and for a distance of 2.5 kpc
or otherwise noted). However, the source was found to be active on 21
January 2000 with a maximum reached luminosity of $\sim 2.5 \times
10^{35}$ \Lunit~(on 2 February 2000).

In Figure~\ref{fig:rxte_lc}{\it b} and {\it c}, close-ups of the count
rate curve are displayed to show the strong variability more
clearly. As can be seen, the source was highly variable on time scales
of days, with episodes of clear activity (luminosities up to
$\sim10^{35}$ \Lunit) followed by episodes of non-activity (with {\it
RXTE}/PCA upper limits on the luminosity of typically a few times
10$^{33}$ \Lunit). In Figure~\ref{fig:rxte_lc}{\it b}, the times of
the {\it BeppoSAX} observations (2000 March 5--6, 8, and 22--23;
Wijnands et al. 2001) are indicated, clearly showing that they were
all taken during intervals when the source was in a non-active
episode. Consequently, during these observations the source was not
detected, with an upper limit on the luminosity of $<10^{32}$
\Lunit~(0.5--5.0 keV). However, only a few days later, on 11 March,
SAX J1808.4--3658 could again be detected with {\it RXTE}/PCA at $\sim
10^{35}$ \Lunit, demonstrating a factor of $>$1000 swing in luminosity
on time scales of days.  The active episodes last typically only a few
days but the non-active episodes could last for weeks to up to
$\sim$44 days (although active episodes with a duration of $<$3 days
might have remained undetected due to the 3--4 day sampling of the
data).  From Figure~\ref{fig:rxte_lc}{\it c}, it is apparent that the
source was also highly variable on time scales of only a few hours. In
particular, the decrease in count rate with a factor of $\sim$20 (from
100 to 5 counts \pers) occurred in $\sim$5 hours and the subsequent
increase took only $\sim$1 hour. The last time SAX J1808.4--3658 could
be detected was on 2000 May 13 in a bulge scan observation. Since then
the source has remained undetectable with the {\it RXTE}/PCA (with
upper limits of typically $10^{33}$ erg s$^{-1}$) and it is presumed
to be in quiescence.

During the 2000 outburst, SAX J1808.4--3658 was spectrally very
constant with a power-law photon index varying between 1.9 and 2.4 (a
mean of 2.16). No correlations were found between the spectral
parameters and time, count rate, or with the episodes of the violent
flaring.  As expected, the count rates were nicely correlated with the
X-ray fluxes and the count rates shown in Figure~\ref{fig:rxte_lc} can
be converted to fluxes using a linear relationship of: flux
($10^{-10}$ \funit; 3--25 keV) = 0.12\pp0.01 $\times$ count rate
(counts \pers~PCU$^{-1}$; 2--60 keV). Because the errors on the count
rates are significantly smaller than those on the obtained fluxes, we
only show the count rate curve.  Furthermore, the {\it RXTE}/PCA bulge
scans resulted only in count rate determinations of SAX J1808.4--3658.
Due to the very brief exposures on the source and its very low count
rates ($< 30$ counts \pers~PCU$^{-1}$ after background subtraction),
the data did not allow to fit the spectral data to obtain the spectral
parameters.  However, because the source spectra did not change
significantly during all our TOO observations, it is likely that also
during the bulge scan observations, the source had a similar power-law
shaped X-ray spectrum.

\section{Discussion}

We presented the {\it RXTE}/PCA light curve of the only known
millisecond X-ray pulsar during its 2000 outburst. The maximum
observed luminosity was only $\sim 2.5 \times 10^{35}$
\Lunit~(assuming a distance of 2.5 kpc), which is about a factor of 10
lower than the luminosities observed during the 1996 and 1998
outbursts. However, due to solar constraints the source could not be
observed with {\it RXTE} before 2000 January 21 so the exact moment of
the start of the outburst is unknown and the peak luminosity could
have been significantly higher. The source exhibited strong
variability by a factor of $\sim$20 on time scales of 1--5 hours, but
by even larger factors on time scales of days. Therefore, the source
became frequently too dim to be detected with {\it RXTE}. The last
detection with {\it RXTE} was on 2000 May 13, almost four months after
the source was first detected.  SAX J1808.4--3658 was also observed
(on March 5--6, 8, and 22--23) during its 2000 outburst with {\it
BeppoSAX}, but during these observations it could not be detected,
with an upper limit on its luminosity (0.5--5 keV) of $\sim10^{32}$
\Lunit~(Wijnands et al. 2001). However, on March 11 and 30, we
detected the source again in the bulge scan observations, giving a
luminosity of about $\sim10^{35}$ \Lunit~(3--25 keV). This
demonstrates that the source was variable with a factor of about
thousand on time scales of only a few days.

The erratic 2000 outburst behavior of SAX J1808.4--3658 raises the
question whether this was an unique event, or if this behavior is more
common for this source. By closely examining the data obtained for the
previous outbursts, we might be able to provide some answers. The 1996
outburst was only observed with the {\it BeppoSAX}/WFC.  The recent
detection of a third type-I X-ray burst with the WFC, $\sim$30 days
after the peak of the 1996 outburst, clearly demonstrates that
accretion was still occurring in this system several weeks after the
WFC could not detect the persistent flux anymore (with upper limits on
the persistent emission of $\sim2\times10^{35}$ \Lunit; 2--28 keV; in
't Zand et al. 2001). Although this is only a small fraction of the
duration of the 2000 low-activity state, it indicates that the source
could have been accreting considerably longer during this outburst
than previously assumed.

The complete {\it RXTE}/PCA light curve obtained during the 1998
outburst is shown in Cui et al. (1998), from which it is clear that
during the last three observations, the source flux behaved
erratically.  After the luminosity had decreased steadily to
$\sim10^{34}$ \Lunit~(2--30 keV), it briefly increased again a few
days later to $\sim4\times10^{34}$ \Lunit~(during the next to last
observation). During the last observation, the luminosity had
decreased to around $10^{34}$ \Lunit, but the source was still
detected (thus, during all 1998 observations the source was detected)
and the behavior of the source after this last observation is unknown
but it cannot be excluded that the source entered a similar low-level
activity state as we observed during the 2000 outburst.  To conclude,
we cannot rule out that these low-activity states of SAX J1808.4--3658
are a common feature of the behavior of the source. However, to
conclusively answer this question several new outbursts of this source
have to be observed with X-ray instruments with enough sensitivity and
with monitoring capabilities (i.e., the {\it RXTE}/PCA bulge scans;
note that the {\it RXTE} all sky monitor or the {\it BeppoSAX}/WFC
were not sensitive enough to detect this source during its 2000
low-activity state; J. in 't Zand 2001 private communication).

The low-activity state of SAX J1808.4--3658 resembles similar states
observed for several other transients (see \S~\ref{intro}). Because it
is unknown when SAX J1808.4--3658 was first active in the 2000
outburst, it is difficult to distinguish for this source between a
failed outburst (similar to what was observed for Aql X-1; Bradt et
al. 2000) or a low-activity state that occurred after a major
outburst.  Even the duration of the low-activity state of SAX
J1808.4--3658 ($\sim$114 days) is in between the duration of the
failed outburst ($\sim$70--80 days) and the low-activity state after a
major outburst ($\sim$150--160 days) of Aql X-1 (note that the
low-activity states observed for 4U 1608--52 and 4U1630--47 can last
for several years).  The question remains if disk instability models
can explain these low-activity states.  The most recent versions of
the disk instability models take also into account irradiation of the
disk by the central X-ray emitting region (Dubus, Hameury, \& Lasota
2001; see Lasota 2001 for a review). Such models can produce the
exponential decay of the X-ray light curves of certain X-ray
transients, but cannot explain other types of light curves. Other
processes like an irradiation-induced increase of the mass transfer
rate from the companion star might be responsible for the different
type of light curves or for the low-level activity states (see, e.g.,
Kuulkers et al. 1997 for a discussion about this as a possible
explanation for the low-level activity in 4U 1630--47).

It is difficult to understand the dramatic, factor of $> 1000$
luminosity variations we observe in SAX J1808.4--3658 as owing to
similarly dramatic variations in the mass accretion rate over such
short ($\sim$1 day) time scales.  Perhaps more modest variations can
trigger transitions between two significantly different luminosity
states.  One possible mechanism is the centrifugal inhibition of
magnetic accretion expected below a critical accretion rate -- the
so-called ``propeller'' regime (Illarionov \& Sunyaev 1975).  The
onset of such centrifugal inhibition has been invoked to explain the
sharp steepening of the light curve decay at the end of the 1998
outburst of SAX J1808.4--3658 (Gilfanov et al. 1998) as well as the
1997 outburst of the neutron star transient Aql X-1 (Campana et
al. 1998).  In the 1998 outburst of SAX J1808.4--3658, the transition
occurred when the X-ray luminosity dropped below $4\times 10^{35}$
erg~s$^{-1}$ (assuming a 2.5 kpc distance), corresponding to a
critical mass accretion rate of $\dot M_{\rm crit}=3.4\times
10^{-10}\,M_\odot$ yr$^{-1}$.  If the propeller effect is indeed
responsible for the sharp break in the 1998 light curve, then small
but erratic variations around $\dot M_{\rm crit}$ could in principle
give rise to the enormous luminosity swings we observed in 2000.  It
is interesting to note that the peak 2000 luminosity of $2.5\times
10^{35}$ erg s$^{-1}$ is very close to the critical transition value
inferred from the 1998 outburst, consistent with a propeller
interpretation.  Although it is not clear what might cause the
required small variations in the mass accretion rate around $\dot
M_{\rm crit}$, such variations may be characteristic of the outburst
tails in soft X-ray transients.  In that case, the dramatically
variable behavior observed in SAX J1808.4--3658 (but not in other
transients) may be an accident of the system's $\dot M_{\rm crit}$,
which depends upon the neutron star's spin period and magnetic field
strength.

The propeller mechanism might also explain why SAX J1808.4--3658
became as faint as it did in between the active episodes (to quiescent
luminosities; Dotani, Asai, \& Wijnands 2000).  It is unclear whether
or not the non-active episodes during the 2000 outburst in between the
active ones can be regarded as true quiescent states, because of the
presence of an inner accretion disk in the system (down to the
magnetosphere of the system), which is presumably absent in
quiescence. But it is also unclear how (or even if) such an accretion
disk would affect the X-ray properties of the source compared to the
true quiescent state.

\acknowledgments

This work was supported by NASA through Chandra Postdoctoral
Fellowship grant number PF9-10010 awarded by CXC, which is operated by
SAO for NASA under contract NAS8-39073.  MK acknowledges support by
the Netherlands Organization for Scientific Research (NWO). We thank
Jon Miller for carefully reading a previous version of this paper.

\end{document}